\documentclass[aps,showpacs,twocolumn]{revtex4}
\usepackage[dvips]{graphics,graphicx}
\font\boldmi=cmmib10

\def\/{\over}
\def\<{\left\langle}
\def\>{\right\rangle}
\def\({\left(}
\def\){\right)}
\def\[{\left[}
\def\]{\right]}
\def\d{{\rm d}}
\def\D{{\rm D}}
\def\i{{\rm i}}
\def\e{{\rm e}}

\def\bfgamma{\hbox{\boldmi\char13}}

\def\bfrho{\hbox{\boldmi\char26}}

\def\det{\hbox{det}\,}

\begin{document}
\title{Semiclassical propagator of the Wigner function}
\author{Thomas Dittrich,$^{1}$ Carlos Viviescas,$^{2}$
and Luis Sandoval$^{1}$}
\affiliation{$^{1}$Departamento de F\'{\i}sica, Universidad Nacional,
Bogot\'a D.C., Colombia,
$^{2}$Max Planck Institute for the Physics of Complex
Systems, N\"othnitzer Stra\ss e 38, 01187 Dresden, Germany}
\date{\today}
\begin{abstract}
Propagation of the Wigner function is studied on two levels of
semiclassical propagation, one based on the van-Vleck propagator,
the other on phase-space path integration. Leading quantum corrections
to the classical Liouville propagator take the form of a
time-dependent quantum spot. Its oscillatory structure depends
on whether the underlying classical flow is elliptic or hyperbolic. It
can be interpreted as the result of interference of a \emph{pair} of
classical trajectories, indicating how quantum coherences
are to be propagated semiclassically in phase space. The phase-space
path-integral approach allows for a finer resolution of the quantum
spot in terms of Airy functions. 
\end{abstract}
\maketitle
Quantum propagation in phase space has always been in the shadow of
propagation in conventional (position, momentum) representations. Yet
it is superior in various respects, particularly in the semiclassical
realm: It avoids all problems owing to
projection, such as singularities at caustics. Canonical invariance of
all classical quantities involved is manifest. Boundary
conditions are imposed consistently at a single (initial or final)
time, thus removing the so-called root-search problem and allowing for
initial-value representations. Semiclassical approximations to the
quantum-mechanical propagator have predominantly been seeked in the
form of coherent-state path integrals \cite{her84,sep96,bar01}. 
Closely related are Heller's Gaussian wavepacket dynamics \cite{hel75}
and its numerous modifications. By now, a broad choice of phase-space
propagation schemes is available which score very well if compared to
other semiclassical techniques.

Almost all of these developments refer to the propagation of
wavefunctions in some Hilbert space. Less attention has been
paid to the propagation of Wigner and Husimi functions. They live in
\emph{projective} Hilbert space, i.e., represent the density
operator and are bilinear in the wavefunction. Besides their
popularity, they have a crucial virtue in common: An extension to
non-unitary time evolution is immediate. This opens access to a host
of applications that combine complex quantum dynamics, where a
phase-space representation facilitates the comparison to the
corresponding classical motion, with decoherence or dissipation:
quantum optics and quantum chemistry, nanosystems in biophysics and
electronics, quantum measurement and computation. 

By the scales involved, many of them call for
semiclassical approximations. However, only few such studies exist,
for specific systems predominantly in quantum chaos \cite{ber79},
including dissipative systems \cite{dit90,coh91}. By contrast,
Ref.~\cite{gar04} discusses a new method, Wigner-function propagation
analogous to the solution of classical Fokker-Planck equations. 

As a major challenge, any attempt to directly propagate Wigner
functions requires an appropriate treatment of \emph{quantum
coherences}. As early as 1976, Heller \cite{hel76} argued that
the ``dangerous cross terms'', i.e., the off-diagonal elements of the
density matrix in the relevant representation, can give rise to a
complete failure of semiclassical propagation of the Wigner
function. Quantum coherences are reflected in the Wigner function as
``sub-Planckian'' oscillations \cite{zur01}. They plague semiclassical
approximations by their small scale and by propagating along paths
that can deviate by any degree from classical trajectories. 

In this Letter, we point out how Heller's objections are resolved
by considering \emph{pairs} of trajectories as basis of semiclassical
approximations, and present corresponding expressions for the
propagator of the Wigner function. The concept of trajectory pairs has
been introduced in the present context
by Rios and Ozorio de Almeida \cite{rio02,ozo05}, albeit working in a
strongly restricted space of semiclassical Wigner functions. We here
give a general derivation of the propagator, independently of
any initial or final states. 

Moreover, we go beyond the level of approximations based
on stationary phase. Employing a phase-space path-integral
technique, we construct an improved semiclassical
Wigner propagator in terms of Airy functions. It resolves all
singularities and contains the semiclassical approximations based
on trajectory pairs as a limiting case. The interference patterns we
obtain depend, up to scaling, only on the nature of the underlying
classical phase-space flow---elliptic vs.\ hyperbolic---and in this
sense are universal. While living in projective Hilbert space, this
result is superior to Gaussian wave-packet propagation in that it
allows Gaussians to evolve into non-Gaussians. 

In order to fix units and notations, define the
Wigner function corresponding to a density operator $\hat\rho$ as
$W({\bf r}) = \int\d^f q' \exp(-\i{\bf p\cdot q'}/\hbar)
\<{\bf q}+{\bf q}'/2\right|\hat\rho\left|{\bf q}-{\bf q}'/2\>$
where ${\bf r} = ({\bf p},{\bf q})$ is a vector in
$2f$-dimensional phase space. Its time evolution is generated by a
Hamiltonian $\hat H(\hat{\bf p},\hat{\bf q})$ through the equation of
motion $(\partial/\partial t) W({\bf r},t) = \{H({\bf r}),W({\bf
r},t)\}_{\rm Moyal}$, involving the Weyl symbol $H({\bf r})$ of
the Hamiltonian $\hat H$. The Moyal bracket $\{.,.\}_{\rm Moyal}$
\cite{hil84} converges to the Poisson bracket for $\hbar\to 0$. As
this equation of motion is linear, the evolution of the Wigner
function over a finite time can be expressed as an integral kernel, 
$W({\bf r}'',t'') = \int \d^{2f}r'\,G({\bf r}'',t'';{\bf r}',t')
W({\bf r}',t')$,
defining the Wigner propagator $G({\bf r}'',t'';{\bf r}',t')$. For
autonomous Hamiltonians, it
induces a one-dimensional dynamical
group parameterized by $t = t''-t'$ (in what follows, we restrict
ourselves to this case and use $t$ as the only time argument). This
implies, in particular, the initial condition $G({\bf r}'',{\bf r}';0)
= \delta({\bf r'' - r'})$ and the composition (Chapman-Kolmogorov)
equation $G({\bf r}'',{\bf r}',t) = \int \d^{2f}r\, G({\bf r}'',{\bf
  r},t-t') G({\bf r},{\bf r}',t')$. 

\begin{figure}[floatfix]
\centerline{\includegraphics[width=6.5cm,angle=0]{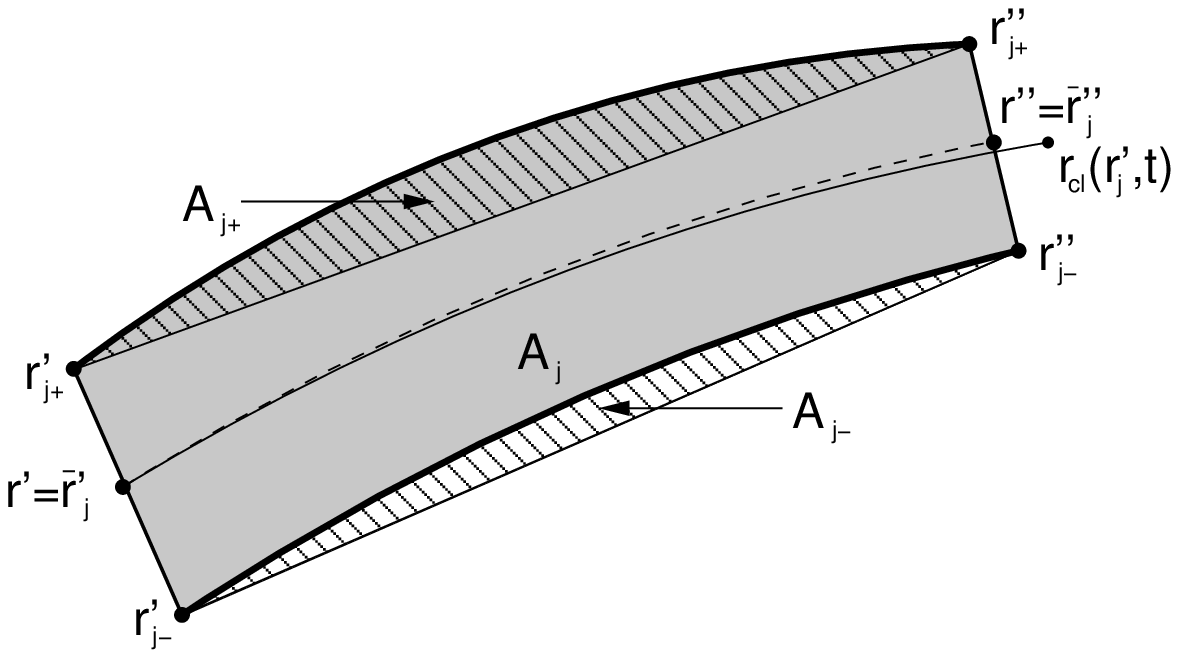}}
\caption{\label{act} 
The reduced action (shaded) of the Wigner propagator in van-Vleck
approximation, Eq.~(\protect\ref{wigvlecks}), is the symplectic area
enclosed between the two classical trajectories ${\bf r}_{j\pm}(t)$
and the two transverse vectors ${\bf r}'_{j+} - {\bf r}'_{j-}$ and
${\bf r}''_{j+} - {\bf r}''_{j-}$ (schematic drawing). The full
central line is the classical trajectory ${\bf r}_{\rm cl}({\bf
r}',t)$, the broken line is the propagation path $\bar{\bf r}_j({\bf
r}',t)$. 
}
\end{figure}

The Wigner propagator can be expressed in terms of the Weyl transform
of the unitary time-evolution operator $U({\bf r},t) =$ $\int\d^f q'
\exp(-\i{\bf p\cdot q'}/\hbar)$ $\<{\bf q}+{\bf q}'/2\right|$ $\hat
U(t)$ $\left|{\bf q}-{\bf q}'/2\>$, called Weyl propagator,
as a convolution,
\begin{equation}\label{wigweyl}
\!\!\!\!\!G({\bf r}'',{\bf r}',t) = \int \d^{2f}R\,
\e^{{-\i\/\hbar}({\bf r}''-{\bf r}')\wedge{\bf R}}
U^*(\tilde{\bf r}_-,t)U(\tilde{\bf r}_+,t),
\end{equation}
with $\tilde{\bf r}_{\pm} \equiv ({\bf r}'+{\bf r}'' \pm
{\bf R})/2$. It serves as a suitable starting point for a
semiclassical approximation, invoking an expression for the Weyl
propagator semiclassically equivalent to the van-Vleck approximation
\cite{ber89,ozo98}, 
\begin{equation}\label{weylvleck}
U({\bf r},t) = 2^f \sum_j
{\exp\(\i S_j({\bf r},t)/\hbar
-\i\mu_j\pi/2\)\/\sqrt{|\det(M_j({\bf r},t)+I)|}}.
\end{equation}
The sum includes all classical trajectories $j$ connecting
phase-space points ${\bf r}'_j$, ${\bf r}''_j$ in time $t$ such that
${\bf r} = \tilde{\bf r}_j \equiv ({\bf r}'_j+{\bf r}''_j)/2$. $M_j$
is the corresponding stability matrix, $\mu_j$ its Maslov
index. The action $S_j({\bf r}_j,t) = A_j({\bf r}_j,t)-H({\bf
  r}_j,t)\,t$, with 
$A_j$, the symplectic area enclosed between the
trajectory and the straight line (chord) connecting ${\bf r}'_j$ and
${\bf r}''_j$ \cite{ber89} (hashed areas $A_{j\pm}$ in
Fig.~\ref{act}). 

\begin{figure}[floatfix]
\centerline{\includegraphics[width=6.5cm,angle=0]{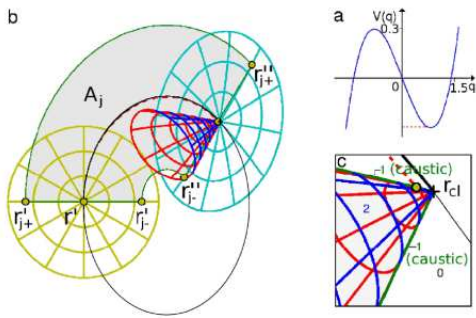}}
\caption{\label{pict} 
Classical building blocks entering the Wigner propagator according to
Eqs.~(\protect\ref{wigvleckg},\protect\ref{wigvlecks}), for a stable  
trajectory starting at ${\bf r}' = (0.636,0)$ near the minimum of the
cubic potential $V(q) = 0.329 q^3-0.69 q$ (panel a). (b) Classical
trajectory ${\bf r}_{\rm cl}({\bf r}',t)$ (black line), a pair of
auxiliary trajectories ${\bf r}_{j\pm}(t)$ (green lines) and
corresponding propagation path $\bar{\bf r}_j({\bf r}',t)$ (red dashed
line). The yellow target pattern is the grid of auxiliary initial
points ${\bf r}'_{j\pm}$ around ${\bf r}'$, parameterized by polar
coordinates. Propagated classically over time $t$, it deforms into the
turquoise pattern around ${\bf r}''$. The red/blue cone is formed by
midpoints $\bar{\bf r}''_j = ({\bf r}''_{j-} + {\bf r}''_{j+})/2$ that
correspond to extrema/saddles of the action. Its boundaries form
caustics separating the region accessed by two midpoints $\bar{\bf
r}''$ from the unaccessible rest.  (c) Enlargement of the area
around ${\bf r}''$, indicating the number of trajectory pairs that
access each region.
}
\end{figure}

Substituting Eq.~(\ref{weylvleck}) in Eq.~(\ref{wigweyl}) leads to a
sum over pairs $j_+$, $j_-$, of trajectories whose respective chord
centers $\tilde{\bf r}_{j\pm}$ are separated by the integration
variable ${\bf R}$. Otherwise, the two trajectories are unrelated. A
coupling between them, as expected on classical grounds, comes about
only upon evaluating the ${\bf R}$-integral by stationary
phase. Stationary points are given by ${\bf r}''-{\bf r}' = ({\bf
r}''_{j-}-{\bf r}'_{j-}+{\bf r}''_{j+}-{\bf r}'_{j+})/2$. Together
with the conditions for the two chords, ${\bf r}'+{\bf r}''\pm{\bf R}
= {\bf r}'_{j\pm}+{\bf r}''_{j\pm}$, this implies 
\begin{equation}\label{midpoints}
{\bf r}' = \bar{\bf r}' \equiv ({\bf r}'_{j-} + {\bf r}'_{j+})/2,\quad
{\bf r}'' = \bar{\bf r}'' \equiv ({\bf r}''_{j-} + {\bf r}''_{j+})/2.
\end{equation}
Stationary points are thus given by pairs of classical
trajectories such that the initial (final) argument of the propagator
is in the middle between their respective initial (final) points
(Fig.~\ref{pict}b). This does \emph{not} require these trajectories to
be identical! They do coincide as long as ${\bf r}'$ and ${\bf r}''$
are  on the same classical trajectory, but bifurcate as ${\bf r}''$
moves off the classical trajectory ${\bf r}_{\rm cl}({\bf r}',t)$
starting at ${\bf r}'$, if the dynamics is not harmonic. 

\begin{figure}[floatfix]
\centerline{\includegraphics[width=3.3cm,angle=0]{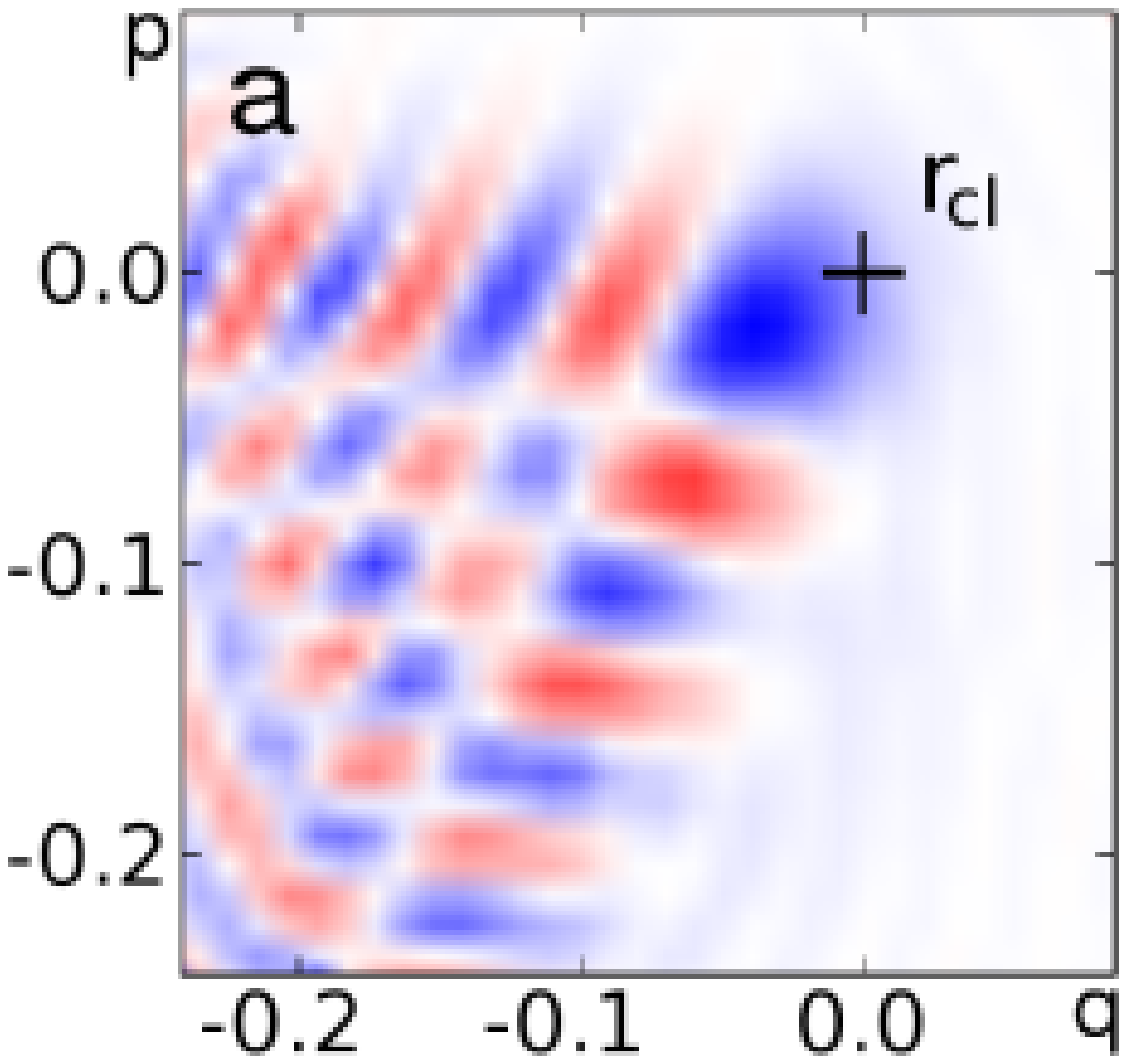}
            \hspace*{-0.2cm}
            \includegraphics[width=3.3cm,angle=0]{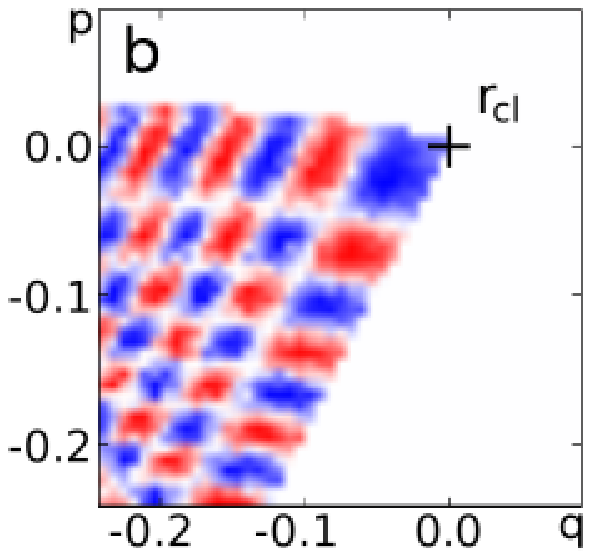}}
\vspace*{-0.2cm}
\centerline{\includegraphics[width=3.3cm,angle=0]{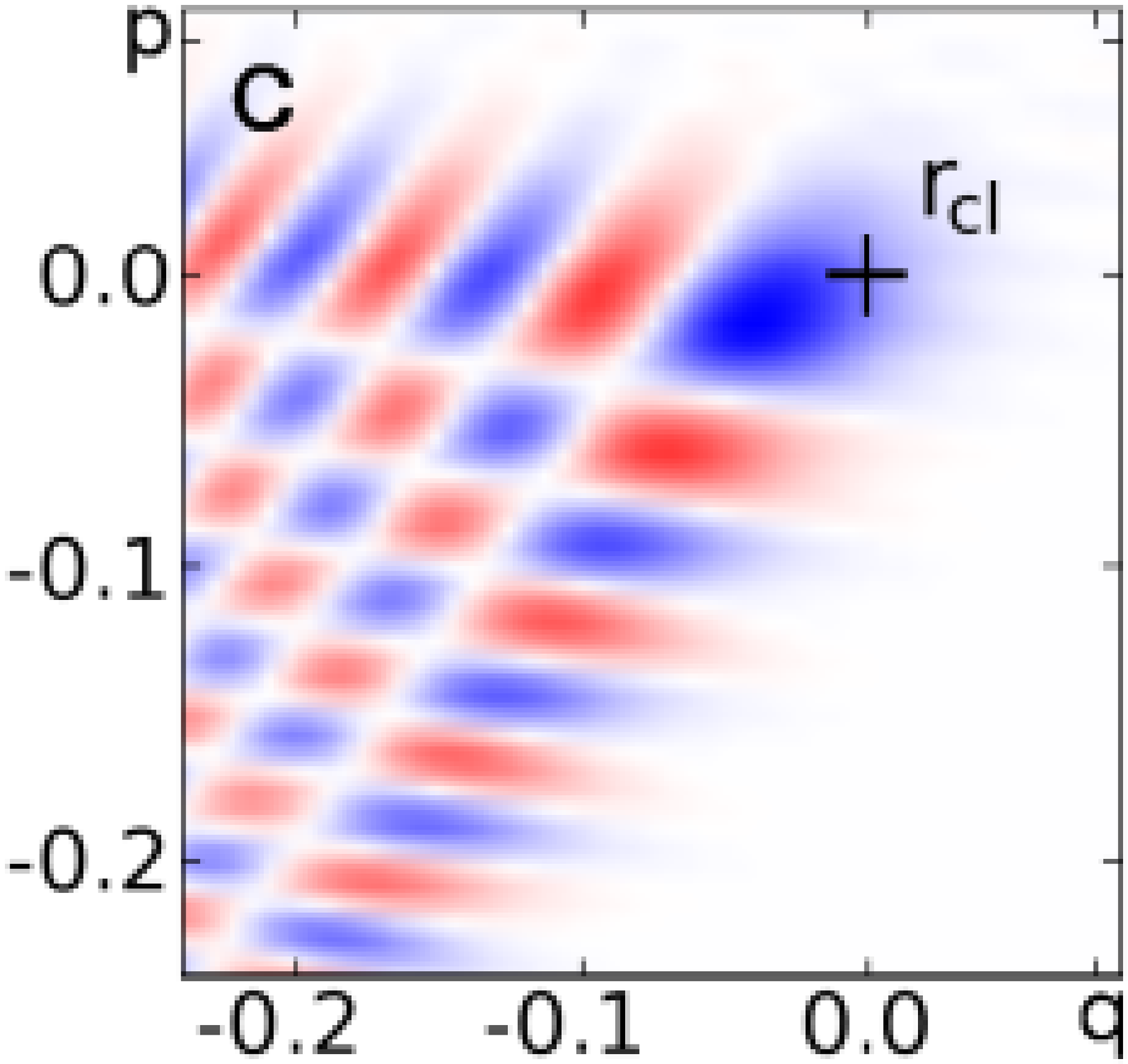}
            \hspace*{-0.2cm}
            \includegraphics[width=3.3cm,angle=0]{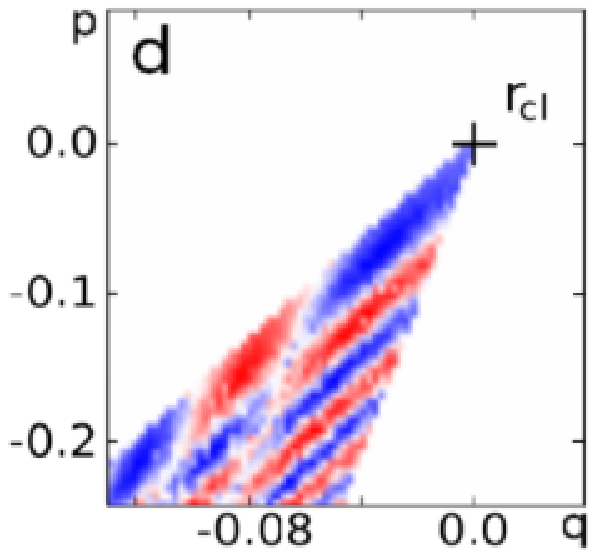}}
\caption{\label{spot} 
Quantum spot replacing the classical delta function on a stable
(elliptic) trajectory near the minimum of a cubic potential as shown
in Fig.~\protect\ref{pict}a, at $t = 1.8$ ($\phi \approx
2\pi/3$). Panel (a) shows the exact quantum result for the Wigner
propagator, (b) and (c) are semiclassical approximations based on
Eqs.~(\protect\ref{wigvleckg}) and (\protect\ref{ajkfgw}),
respectively, all for $\hbar = 0.01$. Frames coincide with that of
Fig.~\protect\ref{pict}c. (d) Quantum spot, according to
Eq.~(\protect\ref{wigvleckg}), for an unstable classical trajectory
near the maximum of the same potential, at $t = 1.0$. Crosses mark the
classical trajectory. Colour code ranges from red (negative) to blue
(positive). 
}
\end{figure}

The resulting semiclassical approximation for the Wig\-ner propagator
is (dot indicating time derivative)
\begin{eqnarray}
G({\bf r}'',{\bf r}',t) &=& {4^f\/h^f}\sum_j
{2\cos\(S_j({\bf r}'',{\bf r}',t)/\hbar-f\pi/2\)\/
\sqrt{|\det(M_{j+}-M_{j-})|}},\label{wigvleckg}\\
S_j({\bf r}'',{\bf r}',t) &=& (\tilde{\bf r}_{j+}-\tilde{\bf r}_{j-})
\wedge({\bf r}''-{\bf r}')+S_{j+}-S_{j-}\nonumber\\
= \int_0^t\d s&&\!\!\!\!\!\!\!\!\!\!\!\!\!
\big[\dot{\bar{\bf r}}_j(s)\wedge
{\bf R}_j(s)-H_{j+}({\bf r}_{j+})+H_{j-}({\bf r}_{j-})\big],
\label{wigvlecks}
\end{eqnarray}
with $\bar{\bf r}_j(t) \equiv ({\bf r}_{j-}(t)+{\bf r}_{j+}(t))/2$,
${\bf R}_j(t) \equiv {\bf r}_{j+}(t)-{\bf r}_{j-}(t)$, and $S_{j\pm}
\equiv A_{j\pm}(\tilde{\bf r}_{j\pm},t)-H_{j\pm}(\tilde{\bf
r}_{j\pm},t)\,t$. The reduced action $A_j = \int_0^t\d s$
$\dot{\bar{\bf r}}_j(s)\wedge {\bf R}_j(s)$ is the
symplectic area enclosed between the two trajectory sections and the
vectors ${\bf r}'_{j+}-{\bf r}'_{j-}$ and ${\bf r}''_{j+}-{\bf
r}''_{j-}$ (Figs.~\ref{act},\ref{pict}b). In general,
Eq.~(\ref{wigvleckg}), as a function of ${\bf r}''$, describes a
distribution that extends from the classical trajectory into the
surrounding phase space, forming a ``quantum spot'' (Fig.~\ref{spot}b)
with a characteristic oscillatory pattern that results from the
interference of the contributing classical trajectories. In
general, it fills only a sector with opening angle $< 2\pi$
(Fig.~\ref{pict}c), where the sum contains two trajectory pairs (four
stationary points). Outside this ``illuminated area'', stationarity
cannot be fulfilled, that is, the ``shadow region'' is not accessible
even for mean paths $\bar{\bf r}_j(t)$. The border is formed by
phase-space caustics along which there is exactly one solution (two
stationary points). As ${\bf r}''$ approaches the classical trajectory
${\bf r}_{\rm cl}({\bf r}',t)$ starting at ${\bf r}'$, from the
illuminated sector, the two solutions $j-$, $j+$ coalesce so that
$M_{j-} \to M_{j+}$, and Eq.~(\ref{wigvleckg}) becomes singular. 
If the potential is purely harmonic, all mean paths coincide with
${\bf r}_{\rm cl}({\bf r}',t)$, and the classical Liouville
propagator, $G({\bf r}'',{\bf r}',t) = \delta({\bf r}'' - {\bf
r}''_{\rm cl}({\bf r}',t))$, is retained. In all other cases,
Eq.~(\ref{wigvleckg}), though based on the van-Vleck propagator,
reflects the structure of stationary points of the action including
third-order terms, with one pair of extrema and one pair of saddle
points. It is formulated in terms of canonically invariant quantities
related to classical trajectories and thus generalizes immediately to
an arbitrary number of degrees of freedom. The propagation of Wigner
functions defined semiclassically in terms of Lagrangian manifolds
\cite{rio02} is contained in Eq.~(\ref{wigvleckg}) as a special case. 

We are now able to resolve Heller's objections \cite{hel76}:
If the two trajectories $j-$, $j+$ are sufficiently separated and the
potential is sufficiently nonlinear, then (i), the propagation path
$\bar{\bf r}({\bf r}',t)$ can differ arbitrarily from ${\bf r}_{\rm
cl}({\bf r}',t)$, and (ii), the phase factor in (\ref{wigvleckg})
exhibits sub-Planckian oscillations. They would couple resonantly to
corresponding features in the initial Wigner function, generating a
similar pattern in the final Wigner function around the endpoint of
the non-classical propagation path. In this way, quantum coherences
are faithfully propagated within a semiclassical approach.

Equations (\ref{wigvleckg},\ref{wigvlecks}) translate into a
straightforward algorithm for the numerical calculation of the
propagator (Fig.~\ref{pict}): (i) Define a local grid (e.g., in polar
coordinates) around the initial argument ${\bf r}'$ of the propagator,
identifying pairs of auxiliary initial points ${\bf r}'_{j-}$, ${\bf
r}'_{j+}$ with ${\bf r}'$ in their middle. (ii) Propagate trajectory
pairs ${\bf r}_{j\pm}(t)$ classically, keeping track of the symplectic
area ${\bf A}_j$ between them. (iii) Find the amplitude and phase
contributed by each trajectory pair and associate them to the final
midpoints $\bar{\bf r}''_j$. They constitute a deformed cone,
projected onto phase space (Fig.~\ref{pict}c). Its ``lower''
(``upper'') surface (red (blue) in Fig.~\ref{pict}c) corresponds to
pairs of extrema (saddles) of the action, respectively: (iv) Superpose
the contributions of the two surfaces, after smoothing amplitude and
phase over midpoints $\bar{\bf r}''_j$ within each of them. 

The caustics in Eq.~(\ref{wigvleckg}) result from applying
stationary-phase integration in a situation where pairs of
stationary points can come arbitrarily close to one another. Since the
underlying van-Vleck propagator admits only up to quadratic terms in
the phase, we seek a superior approach, corresponding to a uniform
approximation. It is available in the form of a path-integral
representation of the Wigner propagator \cite{mar91}, in close analogy
to the Feynman path integral, 
\begin{equation}
G({\bf r}'',{\bf r}',t) =
{1\/h^f}\int\D r\int\D R\,
\e^{-\i S(\{{\bf r}\},\{{\bf R}\})/\hbar}.
\label{pathint}
\end{equation}
Two paths, ${\bf r}(t)$ and ${\bf R}(t)$, have to be integrated over.
The former is subject to boundary conditions ${\bf
r}(0) = {\bf r}'$ and ${\bf r}(t) = {\bf r}''$, the latter is
free. The path action is
\begin{eqnarray}
&&S(\{{\bf r}\},\{{\bf R}\}) = \int_{0}^{t}\d s
\big[\dot{\bf r}(s)\wedge{\bf R}(s)\nonumber\\
&&+H_{\rm W}({\bf r}(s)+{\bf R}(s)/2)-
H_{\rm W}({\bf r}(s)-{\bf R}(s)/2)\big].
\label{pathaction}%
\end{eqnarray}

Equation (\ref{wigvleckg}) is recovered upon evaluating the
path-integral representation in stationary-phase approximation:
Defining ${\bf r}_{\pm}(t) \equiv {\bf r}(t) \pm {\bf R}(t)/2$, with
boundary conditions analogous to Eq.~(\ref{midpoints}), and requiring
stationarity leads to the Hamilton equation of motion for ${\bf
r}_{\pm}(t)$: We again find pairs of classical trajectories that
straddle the propagation path as stationary solutions. 

We will now include cubic terms in the action, with
respect to variations of the path variables. To keep technicalities at
a minimum, we restrict ourselves from now on to a single degree of
freedom and to Hamiltonians of the standard form $H(p,q) = T(p) +
V(q)$, where $T(p) = {p^2/2m}$ while the potential $V(q)$ may contain
nonlinearities of arbitrary order. With this form, $H_{\rm W}({\bf
  r})$ coincides with the Hamiltonian function ``quantized'' by merely
replacing operators with classical variables. As the path integral
readily allows to treat time-dependent potentials, chaotic classical
motion remains within reach. 

Expanding the action (\ref{pathaction}) around ${\bf r}(t) = {\bf
r}_{\rm cl}({\bf r}',t)$ and ${\bf R}(t) \equiv (P(t),Q(t)) = {\bf
0}$, there remain only linear terms in $P$ and linear and cubic terms
in $Q$. Evaluating the ${\bf R}$-sector of the path integral thus
results in an Airy spreading of the propagator, with a rate $\sim
V'''(q_{\rm cl}(t))$, in the $p$-direction. It is superposed
to the classical phase-space flow around the trajectory, i.e.,
rotation (shear) if it is elliptic (hyperbolic). As a consequence, a
spot of the full phase-space dimension develops. Scaling $\bfrho =
(\eta,\xi) = (\mu^{1/4}p,\mu^{-1/4}q)$, with $\mu = T''(p_{\rm
cl})/V''(q_{\rm cl})$, we express the linearized classical
motion as a dimensionless map, 
\begin{equation}\label{contmap}
M(\phi(t)) = \[\begin{array}{cc}\cos\,\phi(t)&-\sin\,\phi(t)\\
\sin\,\phi(t)&\cos\,\phi(t)\end{array}\].
\end{equation}
These maps form a group parameterized by the angle $\phi(t) =
\int_{0}^{t}\d s$ $\sqrt{T''(p^{\rm cl}(s))V''(q^{\rm cl}(s))}$. It
is real (imaginary) if the linearized dynamics is elliptic
(hyperbolic). 

This allows to evaluate also the ${\bf r}$-sector of the path
integral. Transforming the Wigner function to Fourier phase space,
$\widetilde{W}(\bfgamma) \equiv (FW)(\bfgamma) =$ $(2\pi)^{-1}$
$\int\d^2 r$ $\exp(-\i\bfgamma\wedge{\bf r})$ $W({\bf r})$, and the
propagator accordingly, $\tilde{G} = FGF^{-1}$, we obtain ($\bfgamma'
\equiv (\alpha',\beta')$) 
\begin{eqnarray}
&&\!\!\!\!\!\!\!\!\!\!\widetilde{G}(\bfgamma'',\bfgamma',t) =
\delta(\bfgamma''-M(\phi'')\bfgamma')\nonumber\\
&&\!\!\!\!\!\!\!\!\!\!\exp\(-\i({a_{30}\/3}\alpha'^3+
a_{21}\alpha'^2\beta'+a_{12}\alpha'\beta'^2+
{a_{03}\/3}\beta'^3)\).
\label{ajkfgw}
\end{eqnarray}
The coefficients $a_{jk} = \int_{0}^{t}\d s\,\sigma(s)
(\sin\phi(s))^j(\cos\phi(s))^k$ depend on where along the classical
trajectory how much quantum spreading $\sim\sigma(t) = (\mu(t))^{3/4}
\hbar^2 V'''(q_{\rm cl}(t))/8$ is picked up and thus on the
specific system and initial conditions. The Fourier transform from
Eq.~(\ref{ajkfgw}) back to the original Wigner propagator can be done
analytically, after transforming the third-order polynomial in the
phase to a normal form \cite{dit05}. 

The internal structure and the time evolution of the quantum spot
described by Eq.~(\ref{ajkfgw}) are qualitatively different for
elliptic and hyperbolic classical trajectories (real and imaginary
$\phi$, resp.). In the elliptic case, the spot is a periodic function
of $\phi$. In particular, it collapses approximately to a point
whenever $\phi = 2l\pi$, $l$ integer. Close to these nodes, it shrinks
and grows again along a straight line in the $p$-direction, reflecting
the fact that for short time, the quantum Airy spreading $\sim
t^{1/3}$ outweighs the classical rotation $\sim t$. Only sufficiently
far from the nodes, while rotating around the trajectory by
$\phi(t)/2$, the one-d.\ distribution fans out into a two-d.\
interference pattern formed as the overlap of the bright (oscillatory)
sides of two Airy functions, with a sharp maximum on the classical
trajectory (Fig.~\ref{spot}b). In comparison with the corresponding
exact quantum-mechanical result (Fig.~\ref{spot}a) for the quantum
spot, obtained by expanding the propagator in energy eigenstates
\cite{arg05}, the path-integral solution resolves the caustics far
better than Eq.~(\ref{wigvleckg}) (Fig.~\ref{spot}c). 
The hyperbolic case is obtained replacing trigonometric
by the corresponding hyperbolic functions. As a result, along unstable
trajectories there are no periodic recurrences as in the elliptic
case; the spot continues expanding in the unstable and contracting in
the stable direction (Fig.~\ref{spot}d). Isolated unstable periodic
orbits embedded in a chaotic region of phase space exhibit a
degeneracy of the Weyl propagator \cite{ber89}. It allows to account
for scarring in terms of the Wigner propagator \cite{dit05}. 

We have obtained a consistent picture of incipient quantum effects in
the Wigner propagator, both in the van-Vleck approach and in the
path-integral formalism: (i) for anharmonic
potentials, the delta function on the classical trajectory is
replaced by a quantum spot extending into phase space, (ii) its
structure shows a marked time dependence, qualitatively different for
elliptic and hyperbolic dynamics, (iii) it exhibits interference
fringes arising as a product of Airy functions, (iv) it can be
expressed in terms of canonically invariant quantities associated to
pairs of underlying classical trajectories, (v) within each level of
semiclassical approximation used, the propagator retains its
dynamical-group structure. Open issues include: extension to higher
dimensions and to higher-order terms in the action, performance in the
presence of tunneling, application to unstable periodic orbits and
implications for scars, trace formulae, and spectral statistics,
regularization of the ballistic nonlinear $\sigma$-model,
semiclassical propagation of entanglement, and generalization to
non-unitary time evolution. 

We enjoyed discussions with S.~Fishman, F.~Gro\ss mann, 
F.~Haake, H.~J.~Korsch, A.~M.~Ozorio de Almeida, H.~Schanz,
K.~Sch\"onhammer, B.~Segev, T.~H.~Seligman, M.~Sieber, and
U.~Smilansky. Financial support by Colciencias, U.\ Nal.\ de Colombia,
Volkswagen\-Stiftung, and Fundaci\'on Mazda is gratefully
acknowledged. TD thanks for the hospitality extended to him by CIC
(Cuernavaca), Max Planck Institutes in Dresden and G\"ottingen, Inst.\
Theor.\ Phys.\ at Technion (Haifa), Ben-Gurion U.\ of the Negev
(Beer-Sheva), U.\ of Technology Kaiserslautern, and Weizmann Inst.\ of
Sci.\ (Rehovot).


\end{document}